\documentclass[11pt,twoside]{article}
\usepackage{asp2010}

\resetcounters

\begin{document}

\title{Observations of vortex motion in the solar photosphere using HINODE-SP data}
\author{Judith Palacios,$^1$ Laura A. Balmaceda,$^2$ Santiago Vargas Dom\'inguez,$^3$ Iballa~Cabello,$^{1}$ and Vicente Domingo$^{1}$
\affil{$^1$Image Processing Laboratory, Universidad de Valencia, P.O. Box 22085, E-46071, Valencia, Spain}    
\affil{$^2$Inst. de Ciencias Astron\'omicas, de la Tierra y del Espacio - ICATE, CONICET, Av. Espa\~na 1512 Sur, 5400 San Juan, Argentina}
\affil{$^3$ Mullard Space Science Laboratory, University College London, Holmbury St Mary, Dorking, Surrey, RH5 6NT, UK}
}

\begin{abstract}
In this work, we focus in the magnetic evolution of a small region as seen by Hinode-SP during the time interval of about one hour. High-cadence LOS magnetograms and velocity maps were derived, 
allowing the study of different small-scale processes such as the formation/disappearance of bright points accompanying the evolution of an observed convective vortical motion.
\end{abstract}
\section{Introduction}

The study of the small magnetic elements dynamics is fundamental to understand the interaction between magnetism and convection. Solar granulation is the evidence of the hot plasma going upward  
and coming down in the intergranular lanes.  G-band  bright points (thereafter GBPs) are found to be tracers of the convective motion, as they mainly move along intergranular lanes. 
In these movements, GBPs can get close to a strong downflow and start spiralling like water in a bathtub, as reported by \citet{bonet2008}. This spiral motion was also found at larger scales,
 e.g. in supergranular junctions \citep{attie2009}. Moreover, the way of tracking these vortex motions is quite different: computing horizontal velocity fields by using Local Correlation Tracking 
techniques, as one of the first works made about small-scale vortices in the photosphere of \citet{brandt1988}; using either passive tracers, like balltracking \citep{attie2009}  or active tracers,
 as GBPs in \citet{bonet2008}  can be good tools to find these elusive events.

\section{Observations}

The HOP 0014 was a joint campaign using HINODE instruments and the
Swedish~ 1-m Solar Telescope (SST), among other ground-based telescopes at the
Canary Islands Observatories. On September 29, 2007 a continuous sequence of high-quality
data was taken. The target was a quiet Sun region close to disk center
observed at a variety of wavelengths. The description of the full set of
observations can be found at \citet{balmaceda2010}. In this work, we focus
 in the data taken with Hinode Spectro Polarimeter  \citep{ichimoto2008}.  
This data set consist of the full Stokes vector on the 
lines Fe~$\textsc{i}$ 630.15 and 630.25 nm, recorded from 08:20 to 09:45 UT in dynamic mode. The full
FOV was 2.66'' $\times$  40.57'' and scanned with a cadence of $~$35 s. The exposure
time was 1.6 seconds leading to a noise level of  1.4$\times$ $10^{-3}I_{c}$ ($I_{c}$ is the mean continuum intensity) and with spatial resolution of the
SP measurements of 0''15. The standard reduction procedure was applied to the SP data: correction for dark-current,
flat-field and instrumental cross-talk.

\section{Analysis}

\begin{figure}[h!]
\center
\includegraphics[angle=90,scale=0.50]{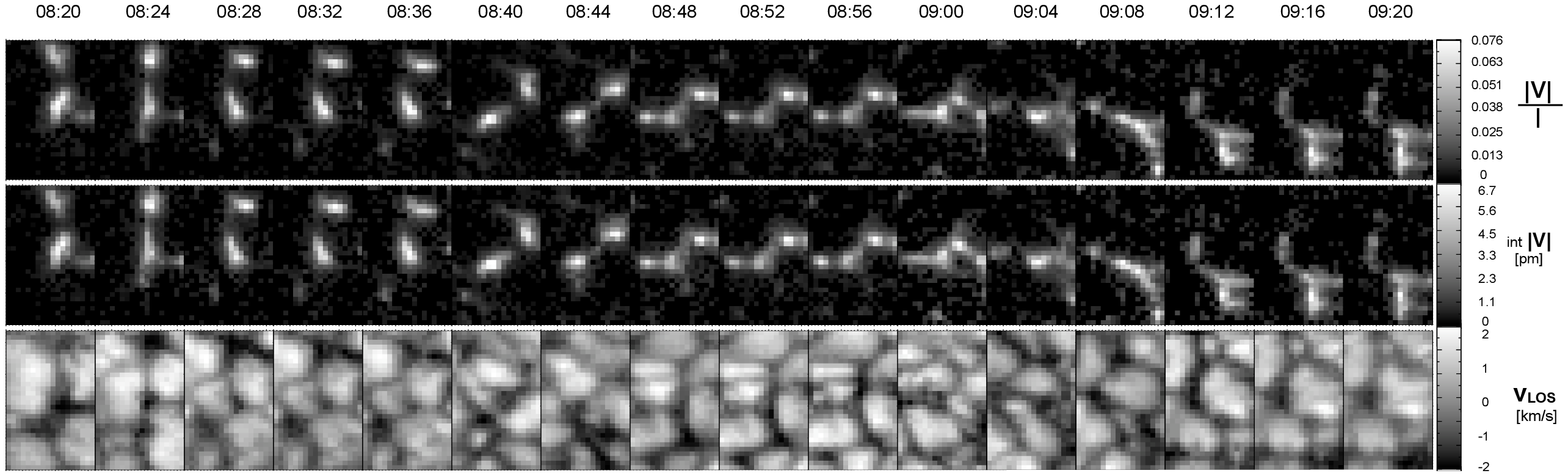}
\caption{From top to bottom rows: sequences of unsigned LOS magnetograms, integrated Stokes $V$ profile and LOS velocity, respectively.}\label{fig1}
\end{figure}

Fig.~\ref{fig1} shows  the SP data sequence. From 08:20 UT to 09:20, images every 4 minutes are presented. The top row displays the LOS magnetograms obtained by computing the max $(V/I)$. 
In the middle sequence the integral of the absolute value of Stokes $V$ is shown, and in the bottom one the line-of-sight (LOS) velocity. 
The magnetic structure, as seen in the magnetograms and the unsigned Stokes $V$ maps, integrated from -29pm to +29pm, shows the same configuration than in the first row, 
and can be generally described as  having two magnetic lobes. At 08:48, some bright points (as seen in the continuum and G-band, though not shown here) start developing between the two lobes, 
and some of them can be seen moving 
nearby the southern lobe (North at the top of the figure). The apparent rotation of these magnetic structure, about 150$^\circ$ in one hour, is observed in this sequence. These magnetic patches 
follow a path whose common center is a draining point. At 09:04, more GBPs develop in the lower right corner and the rotation becomes more evident.\\

The velocity maps were calculated by the center-of-gravity method \citep{uitenbroek2003}. These maps, which follow the granulation behaviour, show in black the redshifted parts, (i.e., downflows), 
and in white the blueshifts, that is, essentially, the granulation structure. The magnetic patches stay most of the time into intergranular lanes. A strong downflow region at 08:44 UT is visible
 there. The $V$ profiles in this region are asymmetric double-lobed, confirming the presence of gradients of velocity or magnetic field \citep{deltoro2003, morinaga2007}. \\

\begin{figure}[h!]
\center
\includegraphics[scale=0.35]{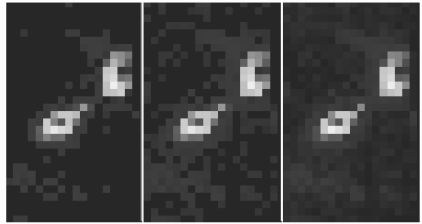}
\caption{From left to right: Spectropolarimetric scan sample image with a 5$\sigma$-mask, 3$\sigma$-mask, and without masking respectively.}\label{fig2}
\end{figure}

Fig.~\ref{fig2} shows a sample of the masks that can be used for removing noise in spectropolarimetric scans. Due to the short exposure time, the noise level increases considerably. 
In order to remove the noise present in the data, we apply different thresholds to create masks. Depending on the purpose, a very restrictive selection can mask out the noisy pixels but 
also remove some small-scale structures.  In Fig.~\ref{fig1} the original image without removing any noise is shown at the right.
 The left and central panel show  5$\sigma$ and 3$\sigma$ masks respectively. In the 5$\sigma$ image, is evident that the main magnetic structure remains unaltered 
as well as some conspicuous bright points (see for instance the BP chain to the left of the northern lobe, or the ones located in the south of the southern lobe). 
However, in the 3$\sigma$ image, we can trace better the aforementioned structures.

\begin{figure}[h!]
 \center
\includegraphics[scale=0.34]{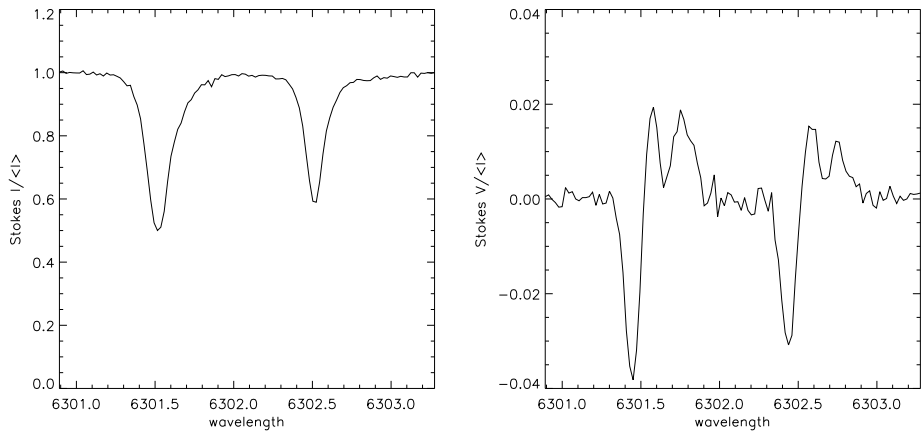}\\
\caption{Left panel: Stokes $I$ profiles. Right panel: abnormal $V$ Stokes profiles.}\label{fig3}
\end{figure}

Looking for downflows in the maps, we found one of the largest asymmetries in the Stokes $V$ profiles that are detected in the SP sequence is the one presented in Fig.~\ref{fig3}. Three-lobed Stokes $V$ profiles are observed in a region with a strong downflow, of about 4 km s$^{-1}$, at 09:21 UT (not shown in Fig.~\ref{fig1}), located in a bright point area, and the \textbar{V}\textbar/I signal becomes more
intense and decreases again along 6 maps (more than 3 minutes), that corresponds to the lifetime of this event. Asymmetries in $V$ profiles can be due to very strong gradients of velocity or magnetic
field, so this behaviour could be the result of a possible convective collapse event. However, there are
other ways of producing such profiles: when two or more components with large Doppler shifts relative to each other overlap \citep{sigwarth2001}. \\

\section{Discussion and concluding remarks}
We have analyzed a sequence of Hinode SP data, applying a 3$\sigma$-mask,
 we justify the use in this work, since a 5$\sigma$ is more restrictive for identifying small magnetic elements. Magnetograms and velocity maps were obtained allowing to
 define the magnetic structure of interest and locate it in the intergranular lanes. The rotating structures become evident in the sequence.
 The appearance of downflows corresponding to BP locations were identified and as well as asymmetric Stokes $V$ profiles in those areas. Further detailed analysis of formation and
disappearance of bright points associated to the observed vortex motion will
be presented in a upcoming paper (Vargas Dom\'inguez  et al., in preparation).

%\section{}   %%% Top level section head (remove "%" symbol)
%\subsection{}   %%% Second level section head (remove "%" symbol)
%\subsubsection{}   %%% Lowest level section head (remove "%" symbol)
%\section*{}    %%% Unnumbered top level section head (remove "%" symbol)
%\subsection*{}   %%% Unnumbered second level section head (remove "%" symbol)
\acknowledgements %%% Text of acknowledgements runs on after this command.
 ~~\emph{Hinode} is a Japanese mission developed and launched by ISAS/JAXA, with NAOJ as domestic partner and NASA and STFC (UK) as international partners. It is operated by these agencies in co-operation with ESA and NSC (Norway).

%%% THE BIBLIOGRAPHY
%%%
%%% CONSULT SECTION 3 OF "INSTRUCTIONS FOR AUTHORS" FOR HOW TO USE NATBIB.
%%% AUTHORS ARE ENCOURAGED TO USE EITHER THE "THEBIBLIOGRAPY" ENVIRONMENT
%%% BY UNCOMMENTING (DELETING THE "%" SYMBOL) THE COMMANDS BELOW, OR BY
%%% USING THE BIBTEX ENVIRONMENT. TO FIND OUT WHICH IS APPLICABLE TO YOUR
%%% CONTRIBUTION, CONSULT THE VOLUME EDITORS FOR YOUR PROCEEDINGS.
%%%

\end{document}